\documentclass{egpubl}
\usepackage{eg2026}
 
\ConferencePaper        % uncomment for (final) Conference Paper
\CGFccby
\usepackage[T1]{fontenc}
\usepackage{dfadobe}  
\biberVersion
\BibtexOrBiblatex
\usepackage[backend=biber,bibstyle=EG,citestyle=alphabetic,backref=true]{biblatex} 
\electronicVersion
\PrintedOrElectronic

\ifpdf \usepackage[pdftex]{graphicx} \pdfcompresslevel=9
\else \usepackage[dvips]{graphicx} \fi

\usepackage{egweblnk} 
\usepackage{duckuments}
\usepackage{amsmath}
\usepackage{booktabs} % For formal tables
\usepackage{float}
\usepackage{layouts}
\usepackage{wrapfig}
\usepackage[ruled,vlined]{algorithm2e}
\usepackage{bbm}
\usepackage{xcolor}
\usepackage{CJKutf8}

\newcommand{\revision}[1]{\textcolor{black}{#1}}

\definecolor{staticColor}{HTML}{005AB5}

\definecolor{dynamicColor}{HTML}{DC3220}

\usepackage{mfirstuc}
\usepackage{amsfonts}
\MFUnocap{et}
\MFUnocap{al}

\newcommand{\refequ}[1] {Eq.~\eqref{eq:#1}}

\newcommand{\reffig}[1] {Fig.~\ref{fig:#1}}

\newcommand{\capFont}[1]{\mathcal{#1}}
\newcommand{\vecFont}[1]{\mathbf{#1}}

\newcommand{\Rnd}[1]{\mathbb{R}^{#1}}

\title[Differentiable Variable Fonts]%
      {Differentiable Variable Fonts}

\author[K. Parikh, D. Kaufman, D. Levin, A. Jacobson]%
{\parbox{\textwidth}{\centering
Kinjal Parikh$^{1, 2}$,\;
Danny M. Kaufman$^{1, 2}$,\;
David I.W. Levin$^{1, 3}$,\;
Alec Jacobson$^{1, 2}$\\[4pt]
$^{1}$University of Toronto,
$^{2}$Adobe Research,
$^{3}$NVIDIA
}}

\begin{document}

% uncomment for using teaser
\teaser{
\vspace{-1cm}
\includegraphics[width=0.925\linewidth]{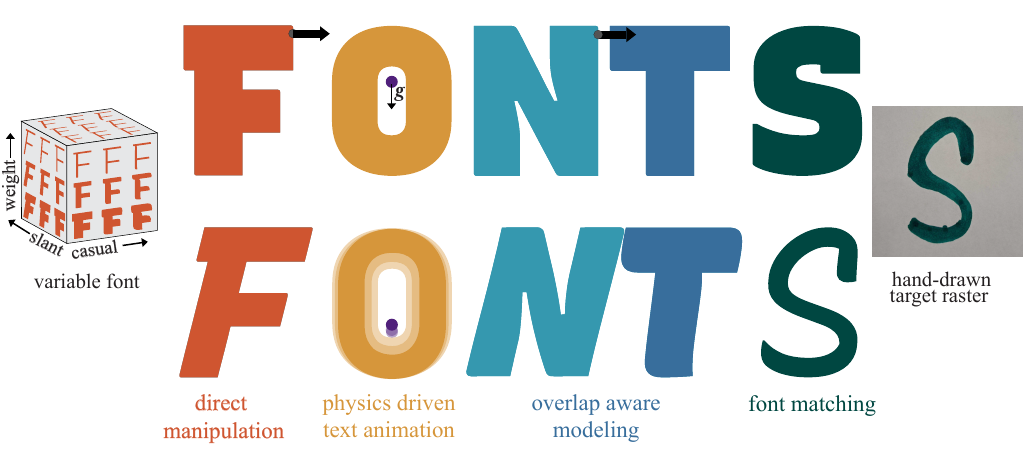}
 \centering
  \caption{We introduce differentiable variable fonts: a mathematical formulation of variable font interpolation made differentiable, enabling gradient-based optimization directly in variable font space. This combines the inherent design guarantees of fonts (legibility, readability, stylistic consistency) with direct differentiable control over text geometry. From left to right: ("F") direct manipulation that adapts glyph shapes while preserving design intent, ("O") physics-driven text animation, ("NT") overlap-aware modeling of glyph interactions, and ("S") font matching with a hand-drawn raster target.}
\label{fig:teaser}
}

\maketitle
%-------------------------------------------------------------------------
\begin{abstract}
      Typography is an essential component of visual communication. 
      Editing and animating text appearance for graphic designs, title sequences, commercials, and logos remain highly skilled tasks requiring detailed, hands-on efforts from pro artists. Automating these challenging manual workflows requires balancing the competing goals of maintaining a text's legibility and aesthetics, while enabling creative expression. Variable fonts, recent parametric extensions to traditional fonts, offer the promise of new ways to ease and automate typographic design and animation. 
      Variable fonts provide custom-constructed parameters along which fonts can be smoothly varied. These parameterizations could then potentially serve as high-value continuous design spaces, opening the door to modern automated design-optimization tools. However, currently variable fonts are underutilized in creative applications, exactly because artists so far still need to manually tune font parameters. Our work provides intuitive and automated font design and animation workflows with differentiable variable fonts. To do so we distill the current variable font specification to a compact mathematical formulation that differentiably connects the highly non-linear, non-invertible mapping of variable font parameters to the underlying vector graphics representing the text. 
      In turn, this enables us to construct a differentiable framework, with respect to variable font parameters, that allows us to perform gradient-based optimization of energies defined on vector graphics control points, and likewise, via differentiable SVG rasterization, on target rasterized images. 
      We demonstrate the utility of this framework with a range of applications, including direct shape manipulation, overlap aware modeling, physics-based text animation, and automated font-design optimization. 
      Our work now enables leveraging the carefully designed affordances of variable fonts with differentiability to use modern design-optimization technologies, and so opens new possibilities for easy, intuitive and expressive typographic design workflows. \\
%-------------------------------------------------------------------------
%  ACM CCS 1998
%  (see https://www.acm.org/publications/computing-classification-system/1998)
% \begin{classification} % according to https://www.acm.org/publications/computing-classification-system/1998
% \CCScat{Computer Graphics}{I.3.3}{Picture/Image Generation}{Line and curve generation}
% \end{classification}
%-------------------------------------------------------------------------
%  ACM CCS 2012
%The tool at \url{http://dl.acm.org/ccs.cfm} can be used to generate
% CCS codes.
%Example:


\end{abstract}   
%-------------------------------------------------------------------------
\section{Introduction}

%introduce; 
Text is a fundamental element of visual communication, used across design, animation, and digital media. 
But editing and animating text remains a challenging task that demands significant time and manual effort from artists. 
% dig a hole
Current workflows for manipulating typography rely on converting text, typed in a fixed font instance, into images, meshes, or SVG paths. 
However, this conversion from text to graphics representations is a one-way process that divorces the geometry from the font, making it hard to retain \emph{legibility, readability, and consistent style} (See \reffig{puppetwarp_comparison}).
This disconnect poses a challenge for developing tools for text manipulation and animation.

\begin{figure}
    \centering
\includegraphics[width=\linewidth] {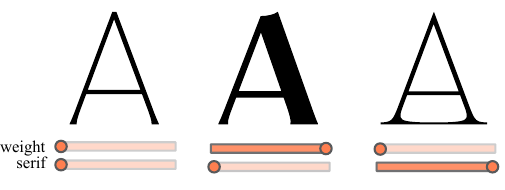}
    \caption{A variable font with weight and serif as variable axes.}
    \label{fig:intro-vfsliders}
\end{figure}

% Fill the hole “In this paper, “
% \alec{This is good. I would scare the reader a bit by foreshadowing the complexity of variable fonts. This will set up their expectations to be appreciative of the differentiable implementation. Talk about the high-dimensionality, the non-linearities, the derivative discontinuities (and even the value discontinuities though not too much since we don't resolve them)}
To address this disconnect between graphics tools and typography, we build on a \revision{recent} font technology: variable fonts. A variable font is a single font file that stores a continuous range of design variants.
Instead of offering a fixed style, a variable font exposes one or more parameters, called \emph{axes}, that control stylistic variation. 
Each axis corresponds to a numerical value that can be adjusted independently and may correspond to conventional typographic properties such as weight or width, or to custom dimensions defined by the type designer (See \reffig{intro-vfsliders}).
This continuous design space is powerful – as long as the text manipulation is constrained to a variable font, it will be legible and consistent in style. 
However, variable fonts are underutilized in creative applications because artists need to manually adjust each parameter in this often high-dimensional space.

% sort of listing contributions
We formalize the \revision{existing OpenType} variable font interpolation mechanism in a compact mathematical model, revealing the highly non-linear, non-invertible mapping of variable font parameters to the underlying vector graphics representing the text (Section \ref{sec:vfmath}).
Based on our formulation, we implement variable font interpolation in a way that exposes gradients with respect to a font's axes, enabling gradient-based optimization directly in the variable font space (Section \ref{sec:optimization}).
This makes it possible to cast a broad range of graphics editing tasks as energy minimization problems defined directly over variable font parameters.

We showcase the utility of our differentiable framework through four applications spanning interactive design, animation, and analysis.
First, we introduce an inverse-kinematics-style interface for direct glyph manipulation, enabling designers to edit shapes via geometric targets rather than abstract axis values (Section \ref{subsec:2dediting}). 
Second we support overlap-aware modeling that automatically resolves collisions (Section \ref{subsec:overlap}), and third a physics-inspired simulation pipeline that animates texts with forces and constraints, while preserving legibility and style. 
(Section \ref{subsec:dynamics}).
Fourth, we demonstrate an image-based font matching optimization, that finds variable font instances that best-capture free-form raster image input targets
(Section \ref{subsec:fontmatching}).

%-------------------------------------------------------------------------

\section{Related Work}

\subsection{Parametric Font Representations}
The idea of fonts as continuous design spaces dates back to early systems like Knuth’s METAFONT \cite{10.5555/892233}, and Adobe’s Multiple Master fonts \cite{AdobeMM5087}, which explored parameterized or interpolated glyph design. 
These efforts laid the groundwork for modern variable fonts \cite{OpenTypeVFSpec}, which extend the OpenType standard to support continuous variation along designer-defined axes.

Alongside industrial developments, academic research has explored parametric representations of fonts—ranging from early work that treated high-level glyph features like stems and serifs \cite{shamir1998feature} or reusable components \cite{hu2001parameterizable} as parameters, to more recent approaches that learn continuous font spaces from examples to support interpolation \cite{campbell2014learning}. 
While these methods can be effective for font generation, they often offer limited user control over the design space — an important requirement in applications like logo design, title sequences, and motion graphics, where maintaining stylistic consistency is critical for brand identity. 
Variable fonts provide a practical alternative: they define a constrained, interpretable space crafted by type designers, allowing controlled manipulation without sacrificing design intent. 
They are also supported across all major browsers and enable efficient optimization, making them well-suited for deployment in real-world design tools.
\revision{Rather than proposing a new font representation, our contribution lies in providing a concise mathematical formalization of the existing OpenType variable font mechanism and making it differentiable, enabling gradient-based optimization.}

\subsection{Font Manipulation}
\begin{figure}
    \centering
    \includegraphics[width=\linewidth]{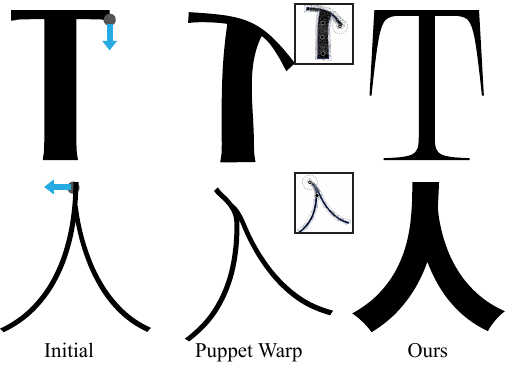}
    \caption{
    General-purpose direct manipulation often compromises legibility and stylistic consistency. 
    We compare with Puppet Warp \cite{liu2014skinning}, default settings with automatically generated pins (see inset).
    For character T, our method maintains style by cleanly extending the serif when the indicated point is dragged downward, whereas Puppet Warp produces distortions. 
    For the Chinese character \begin{CJK}{UTF8}{gbsn}人\end{CJK} (person), Puppet Warp deforms the shape into \begin{CJK}{UTF8}{gbsn}入\end{CJK} (enter), a different character. Our method preserves readability, legibility, stylistic consistency.
    }
    \label{fig:puppetwarp_comparison}
\end{figure}

Digital fonts represent each glyph's geometry as a collection of Bézier curves. 
In standard design workflows, text is converted to vector paths and edited using general-purpose tools such as Adobe Illustrator, Figma, Inkscape, or Affinity Designer. 
This detaches the text from its typographic structure, reducing it to unstructured vector geometry. 
While this provides flexibility, making precise edits requires multiple iterations of careful curve manipulation to preserve legibility and aesthetic appeal. 
The process is time-consuming and demands expert judgment.
Likewise, automatic shape manipulation techniques from geometry processing operate on vector graphics or meshes without regard for typographic structure, and can similarly compromise legibility when applied to text. 
Moreover, many such techniques aim to preserve area, length, or angles under deformation \cite{igarashi2005rigid,sorkine2007rigid,weng20062d,10.1145/1778765.1778775, weber2010controllable, solomon2011killing}, making them ill-suited for common typographic modifications, such as increasing stroke weight or applying slant, which alter area and internal angles in ways these methods are not designed to accommodate. (See \reffig{puppetwarp_comparison})

Attempts to preserve meaningful structure during 2D or 3D shape editing have a long history across various application domains. 
Some methods rely on user-defined constraints \cite{gleicher1992briar, hsu1993skeletal, sutherland1964sketch}, while others attempt to automatically detect and preserve salient structures \cite{cabral2009structure, bernstein2015lillicon, araujo2022locally}. 
In contrast, variable fonts are carefully constructed by type designers to ensure that every point in the design space yields a legible and stylistically coherent instance. 
By allowing the end user to select an appropriate variable font, we can take advantage of this expert-authored structure, with the assurance that edits confined to the font’s defined variation space will preserve legibility. 

\subsection{Kinetic Typography}
Kinetic typography, or text animation, is a well-established motion graphics technique used to capture audience attention \cite{borzyskowski2004animated,kidawara2008kinetic}. The Kinetic Typography Engine \cite{lee2002kinetic} laid the groundwork for modern animation software (e.g., Adobe After Effects and TypeMonkey), where designers manually keyframe properties such as position, scale, rotation, and opacity. For novice users, presets offer one-click animations, though these are often rigid and provide limited creative flexibility.

More recently, image-based approaches have been explored \cite{liu2024dynamic, park2024kinetic, men2019dyntypo}. Wakey-Wakey \cite{xie2023wakey} performs example-based motion transfer by optimizing control points of text vector outlines. However, such methods either neglect core principles of typography, legibility and stylistic consistency, or rely on heuristic regularization, which still cannot guarantee their preservation. In contrast, our work allows animating text directly in the space of variable fonts, where legibility and stylistic coherence are preserved by construction. To drive motion, we employ variational methods for physics based animation.

\subsection{Font Matching}
Font recognition is the task of identifying a typeface given an image of a text fragment. 
It has been studied extensively, with approaches ranging from nearest-class-mean classifiers with learned features \cite{chen2014large} to deep learning–based methods \cite{wang2015deepfont,wang2015real,fontRec2018}. 
These methods aim to select the closest typeface from a large library of static fonts. 
In contrast, our objective is to recover the best-matching instance within a variable font, operating in a continuous parameter space rather than over discrete classes. 
To this end, we combine differentiable rasterization \cite{diffvg} with our differentiable variable font framework to directly optimize font variation axes against a simple image loss without requiring any learning.

%-------------------------------------------------------------------------
\section{Variable Font Mathematics} \label{sec:vfmath}

Variable Fonts\revision{, as defined by the OpenType specification,} are a generalization of earlier efforts like Adobe’s Multiple Master and Apple’s GX fonts. 
Their ability to easily interpolate between designer-specified glyphs belies complex underlying interpolation modes expressed through a combination of pseudocode and descriptive text, distributed across several sections of the OpenType documentation.
In this section we distill this implementation-skewed schema  to a compact mathematical formulation of variable font interpolation. 
This mathematical foundation enables the development of a differentiable variable font framework, which can then be applied to support new modes of interaction and control.

A font file contains a collection of \textit{glyphs}, the geometric descriptions of characters, along with other information such as layout and rendering instructions.
In modern fonts glyph geometry is stored as vector outlines.  
\begin{wrapfigure}{r}{0.34\linewidth}
   \vspace{-0.5\baselineskip}
   \hspace{-0.5cm}
   \includegraphics[width=\linewidth]{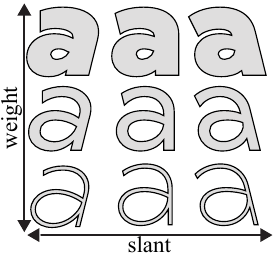}
   \vspace*{-1\baselineskip}
\end{wrapfigure} 
We exactly convert these outlines (including quadratic curves and line segments) to sequences of cubic Bézier segments, so the ordered 2D control-point positions fully determine the shape.
Variable fonts extend this idea by letting glyph outlines vary continuously along one or more design axes(known as \textit{variable axes}).
The inset shows variations of the character ‘a’ from the variable font Afacad Flux, obtained by adjusting its ‘weight’ (wght) and ‘slant’ axes.

\begin{figure}[!h]
\begin{center}
\end{center}
   \vspace{-0\baselineskip}
   \includegraphics[width=\linewidth]{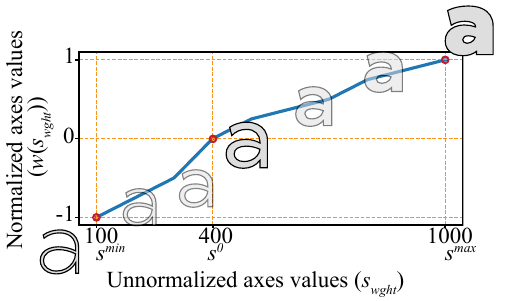}
   \vspace*{\baselineskip}
   \caption{Mapping from user facing slider values to the normalized axes values}
   \label{fig:normalization}
\end{figure}

Each variable axis $i$ is defined over a numeric domain with a minimum, maximum, and default value: 
\( s_i^{min} \leq s_i \leq s_i^{max} \). 
An instance of the variable font is specified by a vector 
\(\vecFont{s} = (s_1, s_2,  \ldots, s_n) \in \Rnd{n} \), where $s_i$ is chosen (by the user via axis sliders in a UI) within the axis’s allowed range.  
Internally, these axis values are normalized to a standard coordinate system via a piecewise-linear transformation, which maps $\{s^\text{min}_i , s^0_i , s^\text{max}_i\}$ to $\{-1,0,1\}$. 
$$
w_i(s_i) = \begin{cases}
  \frac{s_i - s^\text{0}_i}{s^\text{0}_i - s^\text{min}_i} & \text{if } s_i \le s^\text{0}_i \\
  \frac{s_i - s^\text{0}_i}{s^\text{max}_i - s^\text{0}_i} & \text{otherwise } (s_i > s^\text{0}_i),
  \end{cases}
$$ where $s^0_i$ specifies the values of $s_i$ which map to $0$. 
To achieve perceptual uniformity along $s$, the font designer may specify a non-uniform piecewise-linear remapping. 
This remapping must still map ${s^\text{min}_i}$, ${s^0_i}$, ${s^\text{max}_i}$ to $\{-1,0,1\}$ (see \reffig{normalization}).
Throughout the remainder of the paper, we treat 
$\vecFont{w} \in [-1, 1]^n$ as the canonical representation of a font instance, used in all interpolation and optimization computations.

\begin{figure}[!h]
\centering

\begin{minipage}[t]{0.9\linewidth}
    \centering
    \includegraphics[width=0.9\linewidth]{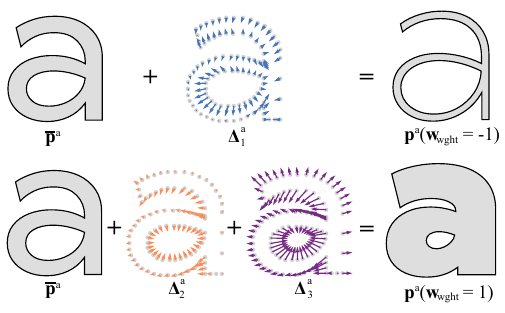}
    
    \vspace{0.5em}
    \small (a) 
\end{minipage}

\vspace{1em}

\begin{minipage}[t]{0.9\linewidth}
    \centering
    \includegraphics[width=0.9\linewidth]{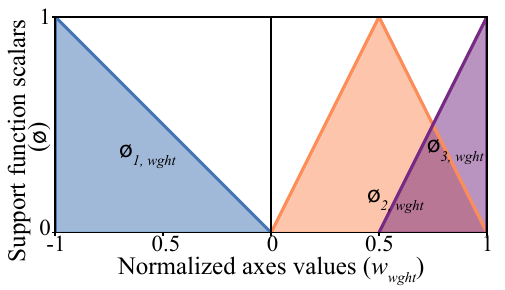}
    
    \vspace{0.5em}
    \small (b) 
\end{minipage}

\caption{
Glyph variations are produced by adding delta sets to the default glyph $\overline{\mathbf{p}}$, with each delta set scaled by a nonlinear support function of $\vecFont{w}$. (a) Example variations of the character ‘a’ from the variable font Afacad Flux at $w_{\text{wght}}=-1$ and $w_{\text{wght}}=0.999$ (other axes at default), showing the contributing delta sets. (b) One-dimensional view of the support functions $\phi$ along the weight axis, indicating the regions of influence of the delta sets in (a).
}
\label{fig:sum_and_support}
\end{figure}

The font explicitly stores the control point locations for the default outline of each glyph, corresponding to $\vecFont{w} = \vecFont{0}$. 
We denote the vector of two-dimensional control points of glyph $g$ in this default configuration as \(\overline{\mathbf{p}}^{\text{ }g} \in \Rnd{k\times 2} \), where $k$ is the total number of control points. 
To obtain glyph variations at $\vecFont{w} \neq \vecFont{0}$, the font provides \emph{delta sets} (displacement vectors) that are appropriately scaled and added to \(\overline{\mathbf{p}}^{\text{ }g} \)
(See \reffig{sum_and_support}(a)). 
We denote the delta set tensor for glyph $g$ as $\Delta^g \in \Rnd{k\times 2 \times m}$, where $m$ is the number of delta sets, and $\Delta^g_{:,:,j}$ is the $j$-th delta set. 

% \begin{figure}[!h]
% \centering
% \begin{subfigure}[t]{0.9\linewidth}
%     \centering
%     \includegraphics[width=0.9\linewidth]{figures/sumofdeltasets.pdf}
%     \caption{}
%     \label{fig:sumofdeltasets}
% \end{subfigure}
% ~
% \begin{subfigure}[t]{0.9\linewidth}
%     \centering
%     \includegraphics[width=0.9\linewidth]{figures/supportfunctions.pdf}
%     \caption{}
%     \label{fig:supportfunction}
% \end{subfigure}    
% \caption{
% Glyph variations are produced by adding delta sets to the default glyph $\overline{\mathbf{p}}$, with each delta set scaled by a nonlinear support function of $\vecFont{w}$. (a) Example variations of the character ‘a’ from the variable font Afacad Flux at $w_{\text{wght}}=-1$ and $w_{\text{wght}}=0.999$ (other axes at default), showing the contributing delta sets. (b) One-dimensional view of the support functions $\phi$ along the weight axis, indicating the regions of influence of the delta sets in (a).
% }
% \end{figure}

Each delta set is active only within a prescribed region of the variation space, and its influence is scaled by a weight function (\emph{support function}).
These weight functions are collected in a matrix $\phi^{m\times n}$, where each row corresponds to a delta set and each column corresponds to a variation axis. 
For the $j$-th delta set, the weight function on axis $i$ is piecewise linear, and is defined by three parameters: the start, peak, and end positions, given by the matrices $\mathbf{W}_{j, i}^{\text{start}}, \mathbf{W}_{j, i}^{\text{peak}}, \mathbf{W}_{j, i}^{\text{end}}$. 
$$
\phi_{j,i}(w_i) = \begin{cases}
  \frac{w_i - W_{j, i}^\text{start}}{W_{j, i}^\text{peak} - W_{j, i}^\text{start}} & \text{if } W_{j, i}^\text{start} \leq w_i < W_{j, i}^\text{peak} \\
  \frac{W_{j, i}^\text{stop} - w_i}{W_{j, i}^\text{stop}  - W_{j, i}^\text{peak}} & \text{if } W_{j, i}^\text{peak} \leq w_i < W_{j, i}^\text{stop} \\
  0 & \text{otherwise}.
\end{cases}
$$
If no support function is specified for an axis, $\phi_{j,i}(w_i)=1$ by default. See \reffig{sum_and_support}(b) for an illustration.

\begin{figure}[!h]
\begin{center}
\end{center}    \includegraphics[width=\linewidth]{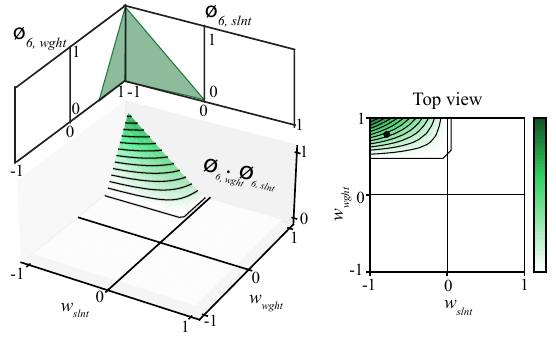}
    \caption{
    The scaling functions of delta sets are highly non-linear and non-invertible as they are defined as products of the support functions across all axes. 
    }
    \label{fig:product-of-supportfunctions}
\end{figure}

\begin{figure}[!h]
\begin{center}
\end{center}    \includegraphics[width=\linewidth]{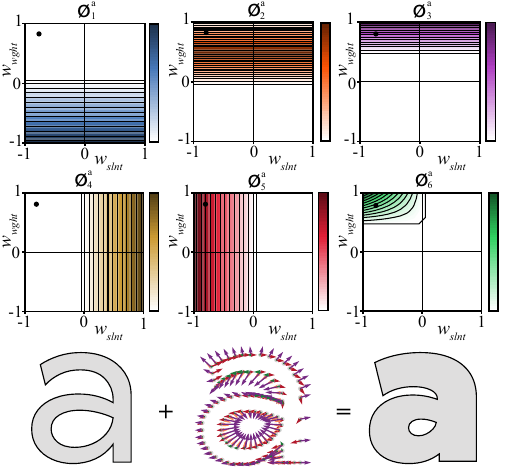}
    \caption{
    Illustration of delta sets and their influence on glyph variation. (Top) Scaling factors for six of the delta sets of the font Afacad Flux, plotted over the two variation axes, weight ($w_{\text{wght}}$) and slant ($w_{\text{slnt}}$). The black dot marks the chosen instance $\vecFont{w}$. (Bottom) The default glyph (left) is deformed by scaled delta displacements (middle) to produce the final glyph at the target instance (right). 
}
    \label{fig:glyph-interpolation}
\end{figure}

The contribution of a delta set is obtained by multiplying its weight functions across all active axes. 
The final glyph shape is then given by the nonlinear interpolation
\begin{equation}
\label{eq:glyph_interpolation}
    \mathbf{p}^g(\mathbf{w}) = \overline{\mathbf{p}}^{\text{ }g} + \Delta^g\cdot\gamma(\vecFont{w})
\end{equation}
\begin{equation}
\label{eq:scaling_factors}
    \gamma(\vecFont{w}) = \prod{\phi(\mathbf{w})},
\end{equation} where $\prod$ is a row-wise serial product and $\cdot$ is tensor contraction with the third dimension of $\Delta^g$. 
We illustrate this process in
\reffig{product-of-supportfunctions} and 
\reffig{glyph-interpolation}. 

\begin{figure}[!h]
\centering    \includegraphics[width=0.7\linewidth]{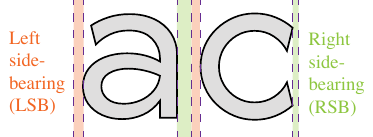}
    \caption{
    Role of Left and Right side bearings in laying out a word. 
    }
    \label{fig:layout}
\end{figure}
\emph{Layout Interpolation.}
In addition to glyph outlines, fonts specify layout quantities that control spacing and alignment in text composition. 
Layout is a complex process handled by text shaping engines, involving features such as kerning, ligatures, contextual alternates, and script-specific behaviors. 
Modeling the entire layout pipeline is beyond the scope of this work. For our purposes, we implement a minimal subset of layout logic sufficient to support word-level text composition. 
Specifically, we model the left and right side bearings.
Side bearings define the space between a glyph’s outline and its bounding box, and together with neighboring glyphs determine inter-glyph spacing (See \reffig{layout}).
These correspond to additional 2D points that define a glyph’s horizontal extent and its spacing relative to neighbors. 
The two points representing the LSB and RSB are interpolated using the same equation as the glyph’s control points (\refequ{glyph_interpolation}).
The final position of the $k$-th control point of the $j$-th glyph in a word of length $l$ is given by 
\begin{equation}
\centering
\label{eq:layout_interpolation}
    \tilde{\mathbf{p}}^l(\Theta) = \sum_{z=0}^{l-1}\left(\overrightarrow{\mathbf{p}}^{\text{ }z}(\Theta_z) - \overleftarrow{\mathbf{p}}^{\text{ }z}(\Theta_z)\right) - \overleftarrow{\mathbf{p}}^{\text{ }l}(\Theta_l) + \mathbf{p}^{\text{ }l}(\Theta_l)
\end{equation}

where \( \Theta = [ \vecFont{w}^1 \xspace \vecFont{w}^2  \xspace \ldots  \xspace\vecFont{w}^l] \) are the parameters for each glyph in the word and $\overleftarrow{\mathbf{p}}^{\text{ }g}$, $\overrightarrow{\mathbf{p}}^{\text{ }g}$ denotes the left and right sidebearing points of $g$ respectively. Note that in this formula, we abuse notation slightly by assuming integer glyph position will index the correct glyph in the font.

%-------------------------------------------------------------------------
\section{Optimization of Variable Fonts}
\label{sec:optimization}

Building on the mathematical formulation in Section~\ref{sec:vfmath}, we now describe a general approach for optimizing variable fonts using automatic differentiation.

We make the mapping from axis weights $\Theta$ to control positions $\vecFont{p}$ in the interpolation pipeline (\refequ{layout_interpolation}) differentiable by implementing it in PyTorch.

A natural concern is that the mapping from axis weights $\Theta$ to control point positions $\vecFont{p}$ is piecewise non-linear and may contain derivative discontinuities at region boundaries. 
However, these occur only on a measure-zero subset of the parameter space, and correspondingly we did not encounter any difficulties, in practice, across all examples presented in this paper.

Our differentiable implementation enables continuous optimization of a wide variety of typography-related energy objectives directly over the variable axis weights. We formalize this as minimizing an energy function defined over axis weights:
\begin{equation}
\label{eq:energy-in-theta} 
    \capFont{E}(\Theta) = \|\capFont{F}(\vecFont{p}(\Theta))\|^2 
\end{equation}
Here, $\capFont{F}$ denotes a differentiable energy evaluated on the interpolated control points $\vecFont{p}(\Theta)$.

\begin{wrapfigure}{r}{0.34\linewidth}
   \vspace{-1\baselineskip}
   \hspace{-0.5cm} \includegraphics[width=\linewidth]{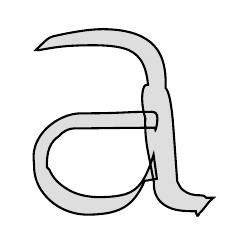}
   \vspace*{-1\baselineskip}
\end{wrapfigure} 
\emph{Choice of variable.} 
The scaling factors of the delta sets ($\gamma(\Theta)$) are highly nonlinear functions due to the product of $\phi(\Theta)$ over all variable axes, and the piecewise-linear structure of $\phi(\Theta)$. 
It may be tempting to consider optimizing for the scaling factors directly, to bypass this nonlinearity. 
However, the $m$-dimensional space of scaling factors is not aligned with the design space of the font. Arbitrary configurations typically fall outside the axis-defined manifold and produce illegible glyphs (see inset).
Moreover, projecting arbitrary scaling factors onto a consistent axis configuration is non-trivial. By contrast, optimizing directly in axis space guarantees that all interpolated instances remain within the design space prescribed by the font.

\emph{Axis limit constraints.} 
The axis weights $\Theta \in [-1,1]^{n \cdot l}$ must remain within their valid bounds, since interpolation is undefined outside these limits. 
In practice, we clamp any out-of-range values before evaluating the interpolation to ensure it is well-defined. 
However, clamping alone creates regions with zero gradient. To address this, we project $\Theta$ back to the valid bounds after each update.

\emph{Optimization strategy.} The optimization framework described above is compatible with a variety of standard gradient-based solvers. In our applications (Section~\ref{sec:applications}), we employ Levenberg–Marquardt for energies defined over sampled outline points, and Adam for image-based energies defined on differentiably rasterized glyphs.

\emph{Preprocessing.} Realizing the variable font interpolation pipeline in practice requires substantial preprocessing. 
The data structures described in Sec. \ref{sec:vfmath} are encoded in compressed lookup tables in the font file, often omitting values implicitly to reduce size—for example, through implied on-curve points, sparsely defined delta sets, default assumptions in region definitions, and inconsistent curve orientations. 
We preprocess these tables into uniform and unpacked representations that expose all geometry and layout data explicitly.

%------------------------------------------------------------------------
\section{Applications} \label{sec:applications}

We now demonstrate the utility of our differentiable variable font optimization framework through several applications. 
These include interactive editing tools (Section \ref{subsec:2dediting}, \ref{subsec:overlap}), 
physics driven kinetic typography (Section \ref{subsec:dynamics}),
and automatic font matching (Section \ref{subsec:fontmatching}). 
Each of these applications builds on the same underlying formulation but defines different energy objectives tailored to the task. 
\revision{We also report representative runtime statistics for these applications in Section 5.5.}

\subsection{Direct Manipulation}
\label{subsec:2dediting}

Current software support for variable fonts is limited to exposing font axes as sliders. 
This interface contrasts sharply with the vector-graphics workflows familiar to digital artists, where geometry is manipulated directly in 2D space. 
Although variable fonts are internally represented as vector graphics, the nonlinear mapping from axis weights to control point positions prevents such direct editing. 
Our differentiable formulation bridges this gap: by expressing editing operations as losses on control points and backpropagating them, we can optimize the axis weights to realize direct manipulation in the variable font domain.

\begin{figure}[!h]
\centering
\includegraphics[width=0.9\linewidth]{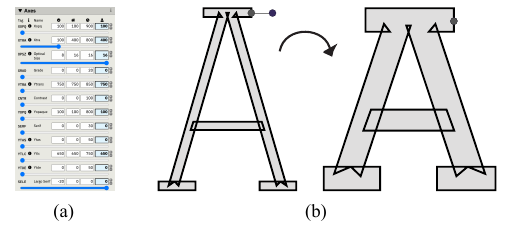} 
\caption{(a) The interface to variable fonts in existing systems exposes the variable font axes as sliders. (b) Our method enables manipulating the font instance by selecting a point on the text outline and dragging it. We optimize the variable font axes weights to match the user intent by leveraging our differentiable implementation. As opposed to typical slider-based workflow, our method enable a more direct and intuitive editing process.}
\label{fig:Simple-2D-interaction}
\end{figure}

To enable direct manipulation, we built an interactive system in which a user can select any point on the outline of a glyph and drag it to a new location. When a user clicks near the outline, the system identifies the closest curve segment and determines the corresponding Bézier parameter $t$. Once selected, the point can be dragged, with the system continuously updating the underlying axes weights to match the target position (See \reffig{Simple-2D-interaction}).

\begin{figure}[!h]
\centering

\begin{minipage}[t]{\linewidth}
    \centering
    \includegraphics[width=0.9\linewidth]{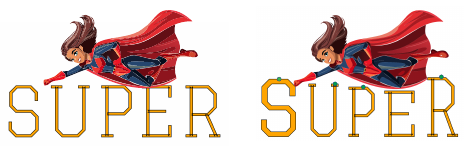}
    \vspace{0.5em}
    
    (a)
    \label{fig:super}
\end{minipage}

\vspace{1em}

\begin{minipage}[t]{\linewidth}
    \centering
    \includegraphics[width=0.9\linewidth]{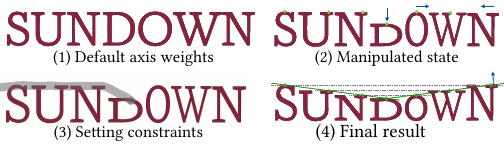}
    \vspace{0.5em}
    
    (b)
    \label{fig:sundown}
\end{minipage}

\vspace{1em}

\begin{minipage}[t]{\linewidth}
    \centering
    \includegraphics[width=0.9\linewidth]{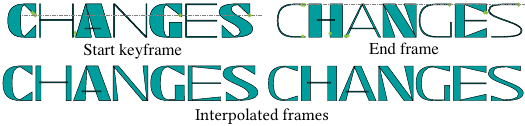}
    \vspace{0.5em}
    
    (c)
    \label{fig:changes}
\end{minipage}

\caption{Examples of our 2D-control based font editing in various settings. (a) shows the result of a sequence of edits made by dragging multiple text outline samples to fit under a background object. 
(b) and (c) show result of constrained edits. 
In (b) outline samples are constrained to be collinear or to form a horizontal line for quickly creating the desired output. 
In (c) constrained edit is used to create keyframes for animation with consistent serif style. 
}
\label{fig:directmanipulation}
\end{figure}

For direct manipulation, we formulate the energy as a least-squares objective that minimizes the difference between the dragged sample point on the outline and its target position.
If $\vecFont{p}(\Theta, t)$ denotes the curve position at parameter $t$ under axis weights $\Theta$, and $x^\ast$ is the user-specified target, the objective is 
\[
    \capFont{E}(\Theta) = \|\vecFont{p}(\Theta, t) - x^\ast\|^2  + \lambda \|\Theta - \Theta_{prev} \|^2,
\]
where the second term regularizes the solution to remain close to the previous font instance.
Additional editing constraints, such as fixing points, enforcing shared $x$ or $y$ coordinates, or encouraging or maintaining collinearity, can be incorporated naturally as extra least-squares terms in the same formulation (See \reffig{directmanipulation}\revision{, video 2:35–2:57}). We solve this optimization using the Levenberg–Marquardt algorithm. We conducted a pilot user study to obtain preliminary feedback on the usability of direct manipulation compared to axis sliders; details are provided in Appendix~\ref{app:userstudy}.

% \begin{figure}[!h]
% \centering
% \begin{subfigure}[t]{\linewidth}
%     \centering
%     \includegraphics[width=0.9\linewidth]{figures/super-girl-result.pdf}
%     \caption{}
%     \label{fig:super}
% \end{subfigure}
% \begin{subfigure}[t]{\linewidth}
%     \centering
%     \includegraphics[width=0.9\linewidth]{figures/sundown-SVGS-FINAL.pdf}
%     \caption{}
%     \label{fig:sundown}
% \end{subfigure}    
% \begin{subfigure}[t]{\linewidth}
%     \centering    \includegraphics[width=0.9\linewidth]{figures/chages-keyframe.pdf}
%     \caption{}
%     \label{fig:changes}
% \end{subfigure}  
% \caption{Examples of our 2D-control based font editing in various settings. 
% (a) shows the result of a sequence of edits made by dragging multiple text outline samples to fit under a background object. 
% (b) and (c) show result of constrained edits. 
% In (b) outline samples are constrained to be collinear or to form a horizontal line for quickly creating the desired output. 
% In (c) constrained edit is used to create keyframes for animation with consistent serif style. 
% }
% \label{fig:directmanipulation}
% \end{figure}

% \label{fig:Simple-2D-interaction}
% \end{figure}

\subsection{Overlap Aware Modeling}
When manipulating typography in graphic design, unintended overlaps can occur between glyphs or with background elements. 
A naive solution is to translate or resize the text, but this often compromises the overall design. 
Variable fonts provide additional degrees of freedom through their axis weights, but manually resolving collisions during iterative modeling can be tedious and time-consuming. 

\begin{figure}[!h]
\centering
\includegraphics[]{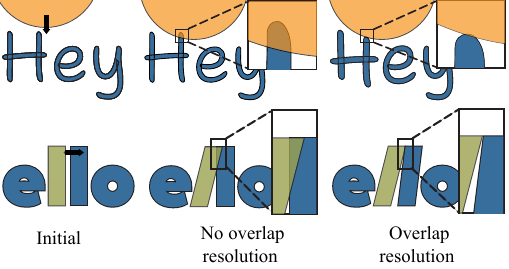} 
\caption{Text overlapping with background objects (top) or itself (bottom) is undesirable in many typography applications. 
Our method allows us to detect and resolve collisions in response to user edits. 
Editing the initial state (left) results in overlaps (center) which are avoided with collision response turned on (right). 
}
\label{fig:overlap}
\end{figure}

Our differentiable framework automates this process by updating the axis weights to resolve overlaps efficiently. (See \reffig{overlap}).
Collision detection is performed by densely sampling points along all glyph outlines and any background objects. 
For each colliding point, we find the closest point on the other surface and its normal, defining a unilateral collision plane. 
Let $\vecFont{t}$ contain the curve indices and Bézier parameters for all points that need to be projected, and let $\vecFont{p}(\Theta, \vecFont{t})$ give the corresponding stacked positions under the current axis weights. 
Similarly, let $\vecFont{b}$ and $\vecFont{N}$ be the stacked closest points and normals. 
We define the collision energy as a differentiable 
%\danny{not everywhere differentiable - right?} least-squares objective:
\begin{equation}
\label{eq:collision-energy}
\mathcal{E}_{\mathrm{collision}}(\Theta) = 
\big\| \max\big(0, \, \vecFont{N}^T(\vecFont{p}(\Theta, \vecFont{t}) - \vecFont{b}) \big) \big\|^2
\end{equation}
which penalizes points that penetrate other surfaces while leaving non-penetrating points unconstrained. 
We iteratively minimize this energy with Levenberg--Marquardt to find axis weights that resolve these collisions.

\label{subsec:overlap}

\subsection{Physics Driven Kinetic Typography}
\label{subsec:dynamics}
Kinetic typography refers to animated text where motion is used to convey emphasis, mood, or narrative. 
It is widely used in graphic design, film, and advertising, but creating compelling motion for typography is difficult: designers must carefully balance visual dynamics with legibility and stylistic consistency. 
Variable fonts provide a natural design space for kinetic typography since their axes allow controlled shape variations without compromising aesthetics or readability. 

\begin{figure}
\centering

\begin{minipage}[t]{\linewidth}
    \centering
    \includegraphics[width=0.9\linewidth]{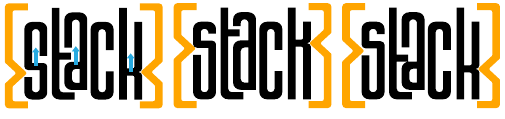}
    \vspace{0.5em}
    
    (a)
    \label{fig:stack}
\end{minipage}

\vspace{1em}

\begin{minipage}[t]{\linewidth}
    \centering
    \includegraphics[width=0.9\linewidth]{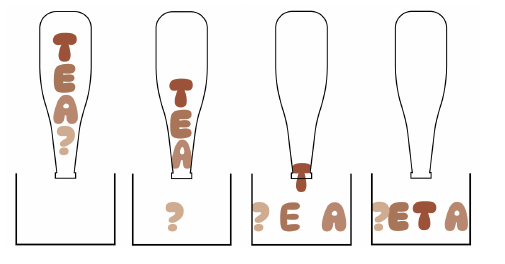}
    \vspace{0.5em}
    
    (b)
    \label{fig:tea}
\end{minipage}

\vspace{1em}

\begin{minipage}[t]{\linewidth}
    \centering
    \includegraphics[width=0.9\linewidth]{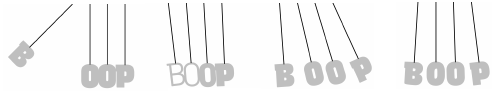}
    \vspace{0.5em}
    
    (c)
    \label{fig:boop}
\end{minipage}

\caption{Examples of kinetic typography animations created with our approach, where text motion is driven by physics-based simulations.
}
\label{fig:animation}

\end{figure}

With our differentiable framework, we can script physics-inspired simulations directly in the space of variable font axes, enabling expressive animations without manually crafting trajectories frame by frame.  
We model momentum, elasticity, and contact in our simulations (\reffig{animation}):
\begin{itemize}
    \item{ Elasticity is expressed as 
\(
\mathcal{E}_{\mathrm{elastic}}(\Theta) = \|\Theta - \Theta_{\mathrm{rest}}\|^2,
\)
which encourages axes weights to remain close to a designated rest configuration. }
% \danny{Stiffness parameter?}}
    \item {Momentum is handled by computing a momentum-predicted state $\Theta_m$ from the current velocity, and introducing a kinetic energy term
\(
\mathcal{E}_{\mathrm{kinetic}}(\Theta) = \|\Theta - \Theta_{m}\|^2,
\)
which drives the solution toward that state. }
%\danny{Density/mass parameter?}}
    \item {Collision handling follows the unilateral constraint formulation described in the Section \ref{subsec:overlap}, \refequ{collision-energy}}
\end{itemize}

For time integration, we adopt an iterative constraint projection approach inspired by projective dynamics \cite{projdyna}. 
Unlike standard projective dynamics, however, we cannot rely on a global linear solve since the collision constraint projection is nonlinear.
Instead, we iteratively minimize the combined elastic, kinetic, and collision energies using the Levenberg--Marquardt optimizer at each timestep.

% \begin{figure}
% \centering
% \begin{subfigure}[t]{\linewidth}
%     \centering
%     \includegraphics[width=0.9\linewidth]{figures/stack.pdf}
%     \caption{}
%     \label{fig:stack}
% \end{subfigure}
% \begin{subfigure}[t]{\linewidth}
%     \centering
%     \includegraphics[width=0.9\linewidth]{figures/tea_new_copy.pdf}
%     \caption{}
%     \label{fig:tea}
% \end{subfigure}    
% \begin{subfigure}[t]{\linewidth}
%     \centering    \includegraphics[width=0.9\linewidth]{figures/boop.pdf}
%     \caption{}
%     \label{fig:boop}
% \end{subfigure}  
% \caption{Examples of kinetic typography animations created with our approach, where text motion is driven by physics-based simulations.
% }
% \label{fig:animation}
% \end{figure}

\subsection{Font Matching}
\label{subsec:fontmatching}
A recurring task in graphic design and document editing is to reproduce the appearance of text found in an image. 
For example, a designer may wish to insert new text into a poster while matching the typography of existing content, or a digitization system may need to identify the closest font instance corresponding to scanned text. 
In all of these cases, the goal is not merely to recognize the font family, but to recover the precise appearance of the text in the observed image. 

\begin{figure}
    \centering
    \includegraphics[width=\linewidth]{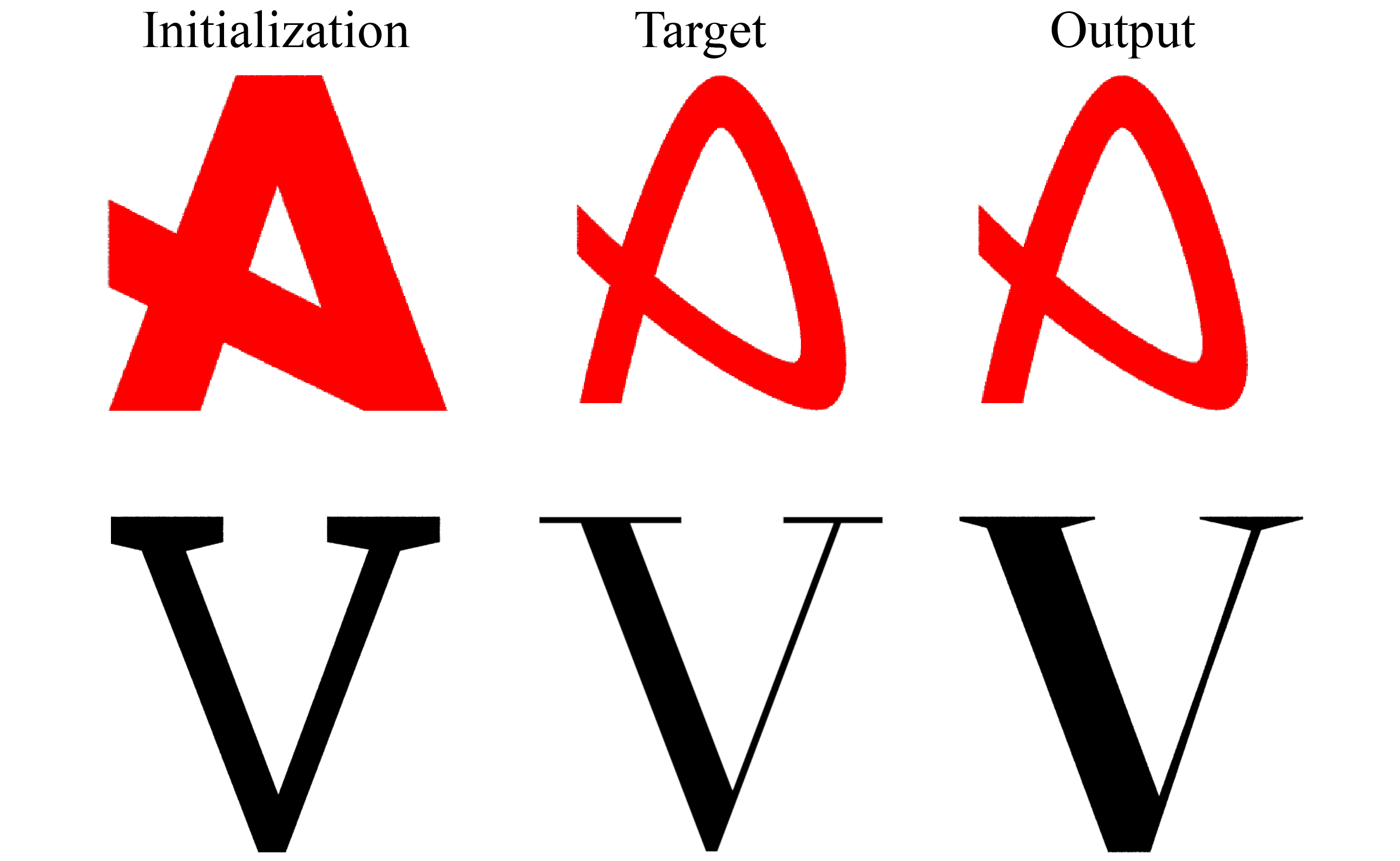}
    \caption{Font matching via differentiable optimization. 
  (Top) A target glyph \textbf{A} from a variable font is reconstructed by optimizing the same font to recover the exact instance. 
  (Bottom) A target glyph \textbf{V} from the non-variable font Didot is approximated using an unrelated variable font (Amstelvar Roman), demonstrating that our method can approximate the target even when the source and target fonts are unrelated.}
    \label{fig:font-matching}
\end{figure}

By combining our differentiable variable font approach with a differentiable SVG rasterizer, we can directly optimize image-based losses. 
Given a target image $I_\mathrm{target}$ and a rendering $I(\Theta)$ from axes weights $\Theta$, we minimize
\[
\mathcal{E}_{\mathrm{image}}(\Theta) = \|I(\Theta) - I_\mathrm{target}\|^2
\]
using gradient descent. 
In practice, we use differentiable vector graphics rasterization \cite{diffvg} and employ an Adam optimizer (with 100 iterations), which reliably converges to visually accurate matches without additional tuning (\reffig{font-matching}). 
\revision{Since glyph geometry often contains overlapping elements, correct rasterization is obtained by specifying the nonzero fill rule in the differentiable SVG rasterizer, without requiring any special handling during optimization.
}
\subsection{Performance}
\revision{We report representative runtime statistics for our applications to provide a qualitative sense of performance. All experiments were run using unoptimized PyTorch code on a MacBook Pro (M1 Pro, CPU only).}

\revision{The average time per update call for interactive direct manipulation, with overlap-aware modeling enabled, is 0.024 s for \reffig{teaser}(F), 0.049 s for \reffig{teaser}(NT), and 0.033 s for \reffig{overlap} (“ello”).}

\revision{Physics-driven kinetic typography is evaluated as an offline simulation. \reffig{teaser}(O) requires 5.27 s for 1000 timesteps, while \reffig{animation}({stack}) takes 9.09 s for 350 timesteps.}

\revision{Font matching is an offline optimization task dominated by differentiable rasterization. Per-iteration runtimes are 0.37 s at resolution 256×256 (\reffig{font-matching}(A)), 1.77 s at 576×512 (\reffig{font-matching}(V)), and 2.21 s at 640×640 (\reffig{teaser}(S)).
}

%-------------------------------------------------------------------------
\section{Limitations and Future Work}

While our framework introduces differentiability into variable font workflows, several important directions remain open.

\emph{Discrete changes.} In addition to continuous variation along design axes, variable fonts can also encode discrete substitutions where the default glyph outline is replaced by a completely different geometry for an assigned part of the design space. 
Our current formulation does not capture these discontinuities, and extending differentiable models to handle such discrete structural changes would broaden the scope of applications.

\emph{Designing variable fonts.} Creating variable fonts today is a labor-intensive process that requires designers to author detailed variation data for every glyph. A promising direction is to invert our framework, using differentiability to assist in generating these displacement sets automatically. Such tools could reduce the burden on type designers by extrapolating full families from a small set of manually designed exemplars.

\emph{Navigating variable fonts.} Variable fonts differ in both the number and semantics of their axes, making it difficult for users to explore and select instances that suit their goals. 
Future work could investigate optimization-based navigation tools, helping users discover font instances that match stylistic, geometric, or functional criteria without exhaustive manual trial and error.

%-------------------------------------------------------------------------

\section{Conclusion}
We introduce a compact mathematical formulation of variable fonts and a differentiable interpolation framework that exposes gradients with respect to font parameters.
This formulation enables a broad class of graphics tasks to be performed directly in variable font space via gradient-based optimization.
Our approach opens up new uses in a wide variety of typography applications, spanning text editing, animation, and analysis.
We believe this technique will catalyze new applications of variable fonts across design and animation, while also inspiring novel authoring tools that reduce the burden of designing them.
To support this vision, we will release our implementation as open-source software, enabling both reproducibility and further exploration.

%-------------------------------------------------------------------------

\section{Acknowledgements}
We thank Skef Iterum for feedback on our variable font formulation, Mengfei Liu for narrating the accompanying video, Zhecheng Wang, Chang Yue, Karran Pandey, Chenxi Liu, and Abhishek Madan for proofreading.
Our research is funded in part by NSERC Discovery (RGPIN–2022–04680), the Ontario Early Research Award program, the Canada Research Chairs Program, a Sloan Research Fellowship, the DSI Catalyst Grant program and gifts by Adobe Inc.

We thank the designers and foundries of the variable fonts used in our figures: Recursive (Fig. 1; Fig. 12(c)), Angst (Fig. 2–3), Noto Sans Simplified Chinese (Fig. 3), Afacad Flux (Fig. 4–8; Fig. 11), Graduate (Fig. 9; Fig. 10(a)), Amstelvar (Fig. 10(b); Fig. 13), Mash (Fig. 10(c)), Shantell Sans (Fig. 11), Kamino (Fig. 12(a)), DynaPuff (Fig. 12(b)), and Movement (Fig. 13).
% bibtex
% \bibliographystyle{sample} 
% \bibliographystyle{eg-alpha-doi}  
% \bibliography{sample}        

% biblatex with biber
\printbibliography 

@techreport{10.5555/892233,
author = {Knuth, Donald E.},
title = {METAFONT: a system for alphabet design},
year = {1979},
publisher = {Stanford University},
address = {Stanford, CA, USA}
}

@techreport{AdobeMM5087,
  author      = {Adobe},
  title       = {Designing Multiple Master Typefaces},
  year        = {1997},
  number      = {5087},
  institution = {Adobe Systems},
  type        = {Technical Note}
}

@manual{OpenTypeVFSpec,
  title        = {OpenType Font Variations Overview},
  author       = {Microsoft},
  year         = {2018},
  organization = {Microsoft},
  url          = {https://learn.microsoft.com/en-us/typography/opentype/spec/otvaroverview}
}

@inproceedings{shamir1998feature,
  title={Feature-based design of fonts using constraints},
  author={Shamir, Ariel and Rappoport, Ari},
  booktitle={International Conference on Raster Imaging and Digital Typography},
  pages={93--108},
  year={1998},
  organization={Springer}
}

@article{hu2001parameterizable,
  title={Parameterizable fonts based on shape components},
  author={Hu, Changyuan and Hersch, Roger D},
  journal={IEEE Computer Graphics and Applications},
  volume={21},
  number={3},
  pages={70--85},
  year={2001},
  publisher={IEEE}
}

@article{campbell2014learning,
  title={Learning a manifold of fonts},
  author={Campbell, Neill DF and Kautz, Jan},
  journal={ACM Transactions on Graphics (ToG)},
  volume={33},
  number={4},
  pages={1--11},
  year={2014},
  publisher={ACM New York, NY, USA}
}

@misc{inkscape,
  author = {{Inkscape}},
  title = {{Inkscape}},
  url = {https://inkscape.org},
  year={2024},
}

@misc{figma,
  author = {{Figma Inc.}},
  title = {{Figma}},
  url = {https://www.figma.com/},
  year={2024},
}

@article{igarashi2005rigid,
  title={As-rigid-as-possible shape manipulation},
  author={Igarashi, Takeo and Moscovich, Tomer and Hughes, John F},
  journal={ACM transactions on Graphics (TOG)},
  volume={24},
  number={3},
  pages={1134--1141},
  year={2005},
  publisher={ACM New York, NY, USA}
}

@inproceedings{sorkine2007rigid,
  title={As-rigid-as-possible surface modeling},
  author={Sorkine, Olga and Alexa, Marc},
  booktitle={Symposium on Geometry processing},
  volume={4},
  pages={109--116},
  year={2007},
  organization={Citeseer}
}

@article{10.1145/1778765.1778775,
author = {Chao, Isaac and Pinkall, Ulrich and Sanan, Patrick and Schr\"{o}der, Peter},
title = {A simple geometric model for elastic deformations},
year = {2010},
issue_date = {July 2010},
publisher = {Association for Computing Machinery},
address = {New York, NY, USA},
volume = {29},
number = {4},
issn = {0730-0301},
url = {https://doi.org/10.1145/1778765.1778775},
doi = {10.1145/1778765.1778775},
journal = {ACM Trans. Graph.},
month = jul,
articleno = {38},
numpages = {6}
}

@article{weng20062d,
  title={2D shape deformation using nonlinear least squares optimization},
  author={Weng, Yanlin and Xu, Weiwei and Wu, Yanchen and Zhou, Kun and Guo, Baining},
  journal={The visual computer},
  volume={22},
  pages={653--660},
  year={2006},
  publisher={Springer}
}

@incollection{weber2010controllable,
  title={Controllable conformal maps for shape deformation and interpolation},
  author={Weber, Ofir and Gotsman, Craig},
  booktitle={ACM SIGGRAPH 2010 papers},
  pages={1--11},
  year={2010}
}

@inproceedings{solomon2011killing,
  title={As-killing-as-possible vector fields for planar deformation},
  author={Solomon, Justin and Ben-Chen, Mirela and Butscher, Adrian and Guibas, Leonidas},
  booktitle={Computer Graphics Forum},
  volume={30},
  number={5},
  pages={1543--1552},
  year={2011},
  organization={Wiley Online Library}
}

@inproceedings{gleicher1992briar,
  title={Briar: A constraint-based drawing program},
  author={Gleicher, Michael},
  booktitle={Proceedings of the SIGCHI conference on Human factors in computing systems},
  pages={661--662},
  year={1992}
}

@inproceedings{hsu1993skeletal,
  title={Skeletal strokes},
  author={Hsu, Siu Chi and Lee, Irene HH and Wiseman, Neil E},
  booktitle={Proceedings of the 6th annual ACM symposium on User interface software and technology},
  pages={197--206},
  year={1993}
}

@inproceedings{sutherland1964sketch,
  title={Sketch pad a man-machine graphical communication system},
  author={Sutherland, Ivan E},
  booktitle={Proceedings of the SHARE design automation workshop},
  pages={6--329},
  year={1964}
}

@inproceedings{cabral2009structure,
  title={Structure-Preserving Reshape for Textured Architectural Scenes},
  author={Cabral, Marcio and Lefebvre, Sylvain and Dachsbacher, Carsten and Drettakis, George},
  booktitle={Computer Graphics Forum},
  volume={28},
  number={2},
  pages={469--480},
  year={2009},
  organization={Wiley Online Library}
}

@article{bernstein2015lillicon,
  title={Lillicon: Using transient widgets to create scale variations of icons},
  author={Bernstein, Gilbert Louis and Li, Wilmot},
  journal={ACM Transactions on Graphics (TOG)},
  volume={34},
  number={4},
  pages={1--11},
  year={2015},
  publisher={ACM New York, NY, USA}
}

@article{araujo2022locally,
  title={As-locally-uniform-as-possible reshaping of vector clip-art},
  author={Ara{\'u}jo, Chrystiano and Vining, Nicholas and Rosales, Enrique and Gori, Giorgio and Sheffer, Alla},
  journal={ACM Transactions on Graphics (TOG)},
  volume={41},
  number={4},
  pages={1--10},
  year={2022},
  publisher={ACM New York, NY, USA}
}

@article{projdyna,
author = {Bouaziz, Sofien and Martin, Sebastian and Liu, Tiantian and Kavan, Ladislav and Pauly, Mark},
title = {Projective dynamics: fusing constraint projections for fast simulation},
year = {2014},
issue_date = {July 2014},
publisher = {Association for Computing Machinery},
address = {New York, NY, USA},
volume = {33},
number = {4},
issn = {0730-0301},
url = {https://doi.org/10.1145/2601097.2601116},
doi = {10.1145/2601097.2601116},
journal = {ACM Trans. Graph.},
month = jul,
articleno = {154},
numpages = {11}
}

@article{liu2014skinning,
  title={Skinning cubic B{\'e}zier splines and Catmull-Clark subdivision surfaces},
  author={Liu, Songrun and Jacobson, Alec and Gingold, Yotam},
  journal={ACM Transactions on Graphics (TOG)},
  volume={33},
  number={6},
  pages={1--9},
  year={2014},
  publisher={ACM New York, NY, USA}
}

@inproceedings{borzyskowski2004animated,
  title={Animated text: More than meets the eye},
  author={Borzyskowski, George},
  booktitle={Beyond the comfort zone: Proceedings of the 21st ASCILITE Conference’, Perth},
  pages={141--144},
  year={2004}
}

@inproceedings{kidawara2008kinetic,
  title={Kinetic typography for ambient displays},
  author={Kidawara, Yutaka and Minakuchi, Mitsuru},
  booktitle={Proceedings of the 2nd international conference on Ubiquitous information management and communication},
  year={2008},
  organization={ACM}
}

@inproceedings{lee2002kinetic,
  title={The kinetic typography engine: an extensible system for animating expressive text},
  author={Lee, Johnny C and Forlizzi, Jodi and Hudson, Scott E},
  booktitle={Proceedings of the 15th annual ACM symposium on User interface software and technology},
  pages={81--90},
  year={2002}
}

@inproceedings{xie2023wakey,
  title={Wakey-wakey: Animate text by mimicking characters in a gif},
  author={Xie, Liwenhan and Zhou, Zhaoyu and Yu, Kerun and Wang, Yun and Qu, Huamin and Chen, Siming},
  booktitle={Proceedings of the 36th Annual ACM Symposium on User Interface Software and Technology},
  pages={1--14},
  year={2023}
}

@article{liu2024dynamic,
  title={Dynamic typography: Bringing text to life via video diffusion prior},
  author={Liu, Zichen and Meng, Yihao and Ouyang, Hao and Yu, Yue and Zhao, Bolin and Cohen-Or, Daniel and Qu, Huamin},
  journal={arXiv preprint arXiv:2404.11614},
  year={2024}
}

@inproceedings{park2024kinetic,
  title={Kinetic typography diffusion model},
  author={Park, Seonmi and Bae, Inhwan and Shin, Seunghyun and Jeon, Hae-Gon},
  booktitle={European Conference on Computer Vision},
  pages={166--185},
  year={2024},
  organization={Springer}
}

@inproceedings{men2019dyntypo,
  title={Dyntypo: Example-based dynamic text effects transfer},
  author={Men, Yifang and Lian, Zhouhui and Tang, Yingmin and Xiao, Jianguo},
  booktitle={Proceedings of the IEEE/CVF Conference on Computer Vision and Pattern Recognition},
  pages={5870--5879},
  year={2019}
}

@inproceedings{chen2014large,
  title={Large-scale visual font recognition},
  author={Chen, Guang and Yang, Jianchao and Jin, Hailin and Brandt, Jonathan and Shechtman, Eli and Agarwala, Aseem and Han, Tony X},
  booktitle={Proceedings of the IEEE Conference on Computer Vision and Pattern Recognition},
  pages={3598--3605},
  year={2014}
}

@inproceedings{wang2015deepfont,
  title={Deepfont: Identify your font from an image},
  author={Wang, Zhangyang and Yang, Jianchao and Jin, Hailin and Shechtman, Eli and Agarwala, Aseem and Brandt, Jonathan and Huang, Thomas S},
  booktitle={Proceedings of the 23rd ACM international conference on Multimedia},
  pages={451--459},
  year={2015}
}

@article{wang2015real,
  title={Real-world font recognition using deep network and domain adaptation},
  author={Wang, Zhangyang and Yang, Jianchao and Jin, Hailin and Shechtman, Eli and Agarwala, Aseem and Brandt, Jonathan and Huang, Thomas S},
  journal={arXiv preprint arXiv:1504.00028},
  year={2015}
}

@InProceedings{fontRec2018,
author="Wang, Yizhi
and Lian, Zhouhui
and Tang, Yingmin
and Xiao, Jianguo",
editor="Schoeffmann, Klaus
and Chalidabhongse, Thanarat H.
and Ngo, Chong Wah
and Aramvith, Supavadee
and O'Connor, Noel E.
and Ho, Yo-Sung
and Gabbouj, Moncef
and Elgammal, Ahmed",
title="Font Recognition in Natural Images via Transfer Learning",
booktitle="MultiMedia Modeling",
year="2018",
publisher="Springer International Publishing",
address="Cham",
pages="229--240",
isbn="978-3-319-73603-7"
}

@article{diffvg,
    title = {Differentiable Vector Graphics Rasterization for Editing and Learning},
    author = {Li, Tzu-Mao and Luk\'{a}\v{c}, Michal and Gharbi Micha\"{e}l and Jonathan Ragan-Kelley},
    journal = {ACM Trans. Graph. (Proc. SIGGRAPH Asia)},
    volume = {39},
    number = {6},
    pages = {193:1--193:15},
    year = {2020}
}

@article{performative_user_study,
author = {Hertzmann, Aaron and Basole, Rahul C. and Ferrise, Francesco},
title = {The Curse of Performative User Studies},
year = {2023},
issue_date = {Nov.-Dec. 2023},
publisher = {IEEE Computer Society Press},
address = {Washington, DC, USA},
volume = {43},
number = {6},
issn = {0272-1716},
url = {https://doi.org/10.1109/MCG.2023.3315759},
doi = {10.1109/MCG.2023.3315759},
journal = {IEEE Comput. Graph. Appl.},
month = nov,
pages = {112–116},
numpages = {5}
}

\appendix

\section{User Study}
\label{app:userstudy}
 To assess whether direct manipulation could be useful for manipulating variable fonts, we conducted a pilot user study with six novice users (5 male, 1 female), all of whom were using variable fonts for the first time. Participants were asked to manipulate a glyph initialized with a variable font’s default parameters to match a target glyph shape sampled from the font’s design space. These tasks were performed using both variable font axis sliders and our direct manipulation interface, which allows users to select, drag, and fix points on the glyph outline.

As a preliminary usability assessment, participants completed the System Usability Scale (SUS) for both interfaces. The direct manipulation interface obtained a mean SUS score of 74.17, compared to 55.41 for the slider-based interface. In an additional comparison question, five out of six participants reported that the direct manipulation interface better captured their intent overall. Qualitative feedback suggested that direct manipulation was perceived as more intuitive and easier to learn, particularly for complex glyphs or fonts with many axes, while sliders were preferred for precise refinement and backtracking.

We emphasize that this was a small-scale pilot study intended to provide preliminary insights rather than statistically conclusive evidence. As discussed in \cite{performative_user_study} many human studies in computer graphics are performative rather than informative; ours is not an exception. Nevertheless, the results provide encouraging evidence that direct manipulation may offer a more intuitive interaction paradigm for variable font editing.
\end{document}